\documentclass[mathleft
]{an}
\usepackage{graphicx}
\usepackage{times}

\begin{document}

\Pagespan{000}{}
\Yearpublication{0000}%
\Yearsubmission{0000}%
\Month{00}%
\Volume{000}%
\Issue{00}%

\title{Techniques for reducing fiber-fed and integral-field spectroscopy data}
   \subtitle{The software package R3D}

   \author{S.F.S\'anchez\fnmsep\thanks{Based on observations collected at the Centro Astron\'mico Hispano Alem\'an (CAHA) at Calar Alto, operated jointly by the Max-Planck Institut f\"ur Astronomie and the Instituto de Astrofísica de Andaluc\'\i a (CSIC)}
          \inst{1}
          }
\titlerunning{Techniques for reducing IFS data}
\authorrunning{S.F.S\'anchez}
\institute{Centro Astron\'omico Hispano Alem\'an, Calar Alto, CSIC-MPG,
  C/Jes\'us Durb\'an Rem\'on 2-2, E-04004 Almeria, Spain
}

\received{}
\accepted{}
\publonline{later}

\keywords{}

\abstract{This paper describes the general characteristics of raw data from
  fiber-fed spectrographs in general and fiber-fed IFUs in particular. The
  different steps of the data reduction are presented, and the techniques used
  to address the unusual characteristics of these data are described in
  detail.  These techniques have been implemented in a specialized software
  package, R3D, developed to reduce fiber-based integral field spectroscopy
  (IFS) data.  The package comprises a set of command-line routines adapted
  for each of these steps, suitable for creating pipelines. The routines have
  been tested against simulations, and against real data from various integral
  field spectrographs (PMAS, PPAK, GMOS, VIMOS and INTEGRAL).  Particular
  attention is paid to the treatment of cross-talk. }

 \maketitle
\section{Introduction}

In fiber-fed spectrographs the slit is not placed at the focal plane of the
telescope. Instead, the light from different locations on the sky is conducted
into the spectrograph using optical fibers.  At one end, the optical fibers
are located in the focal plane of the telescope and at the other they are
rearranged along the slit of the spectrograph, in the so-called pseudo-slit.
Fiber-fed spectrographs are normally used for multi-object spectroscopy (MOS),
where the fiber ends in the focal plane of the telescope can be placed at
different positions on the sky, or for Integral Field Spectroscopy (IFS),
where the fibers are packed in bundles or coupled with arrays of lenses.

Integral Field Spectroscopy (IFS) is a technique for obtaining multiple
spectra simultaneously of a (more or less) contiguous area of the sky.
Although fiber-fed spectrographs are widely used, there are other techniques
to conduct the light from different positions in the sky to the spectrograph,
e.g. lens arrays and image slicers.  These different implementations have
produced a set of instruments, which, while sharing the basics of the
technique, produce very different representations of the spectra at the
detectors. This apparent diversity has led to the development of reduction
techniques and/or packages for each individual instrument (e.g.,
P3d,\cite{beck01}; \cite{roth05}; VIPGI, \cite{vimos05}; GEMINI,
\cite{turn06}). Together with the inherent complexity of this technique, this
has reduced the use of IFS for decades to a handful of specialists, each
usually working with a specific instrument.

IFS developers became aware of this handicap, and have started to produce
standard techniques and tools valid for any integral field unit (IFU). In a
recent effort, the Euro3D RTN (\cite{net02}) has created a standard data
format (\cite{kiss04}), a coding platform (\cite{peco04}) and a visualization
tool (\cite{sanc04}), useful for any of the existing IFU. All these tools are
freely distributed to the community. However most of them are for working with
reduced data, while the reduction itself has still not been addressed with a
generalistic approach.  The astronomical community is confronted with a jungle
of different instrument-specific pipelines, whose procedures and parameters
differ greatly, and whose outputs are difficult to compare.  Building on
earlier work, we present here a summary of the most important steps in the
reduction of data from fiber-fed spectrographs in general, and fiber-fed IFS
in particular. All of them have been implemented in {\tt R3D}, a reduction
package able to reduce data from any fiber-fed IFU, whose main characteristics
we also present.

The structure of this article is as follows. Section \ref{sec:code} describes
the coding platform used to develop {\tt R3D}. Section \ref{sec:raw}
summarizes the major characteristics of the raw data from fiber-fed
spectrographs. The different steps required for a proper data reduction are
explained in Section \ref{sec:red}. In Section \ref{sec:com} we compare {\tt
  R3D} with other packages in existence.  Section \ref{sec:con} summarizes the
conclusions. The data examples shown throughout this article are from PMAS
(\cite{roth05}) in the PPAK mode (\cite{kelz06}), using 2$\times$2 pixel
binning, although {\tt R3D} is not limited to this instrument/setup. For
examples of its use with another instruments, please consult S\'anchez \&
Cardiel (2005).

\section{Coding platform}
\label{sec:code}

We first coded the algorithms in {\tt perl}, using the Perl data language
\footnote{http://pdl.perl.org}, in order to speed-up the algorithm testing
phase. Once we tested the algorithms with simulated and real data, we recoded
the most time consuming ones in {\tt C} to increase their performance. The
{\tt perl} version of {\tt R3D} is fast enough to produce valuable science
frames in a reasonable time, although for instruments with many fibers (like
VIMOS, \cite{lefe03}), or for some slow processes like cross-talk correction
algorithms, we strongly recommend the {\tt C}-version.  Both versions use {\tt
  PGPLOT} as graphical library (Pearson 1995), and the {\tt C} version uses
the Lyon-C-Library and the Euro3D development environment (P\'econtal-Rousset
et al. 2004).  A major advantage of using {\tt perl} and {\tt C} is that they
are freely distributed, platform independent, languages, and therefore {\tt
  R3D} can be installed under almost any architecture without major effort
and at no cost.  {\tt R3D} has been coded as a package of independent
command-line routines, and therefore it is possible to create pipelines and
superscripts that run the same sequence of commands over different input
data. This ensures the homogeneity and repeatability of the data reduction
process.  Download and installation instructions for {\tt R3D} 
are available via its webpage \footnote{http://www.caha.es/sanchez/r3d/}.

\begin{figure}
\centering
   \centering
\resizebox{\hsize}{!}
{\includegraphics[width=\hsize]{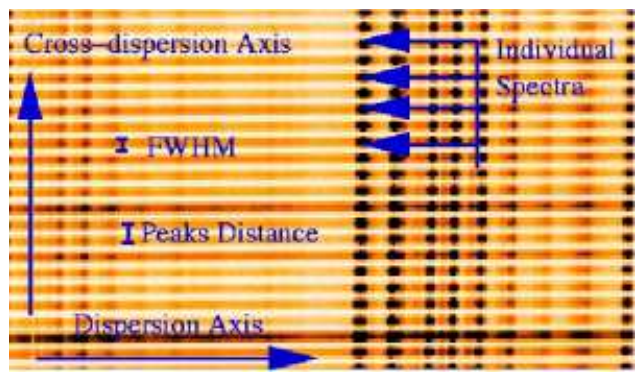}}
\caption{Grayscale of a section of IFS raw data corresponding to PMAS
  in the PPAK mode, in inverse colors (i.e., more intense areas are shown in
  darker colors). Each dark line corresponds to the projection of a spectrum
  along the dispersion axis (indicated with an arrow), which in this case
  corresponds to the X-axis. Spectra are separated $\sim$5 pixels across the
  Y-axis, the cross-dispersion axis, and projected in following a
  pseudo-Gaussian of FWHM$\sim$3 pixels, contaminating the adjacent spectra. }
\label{fig:1}       
\end{figure}

\begin{figure}
  \centering \centering \resizebox{\hsize}{!}
  {\includegraphics[width=\hsize,angle=-90]{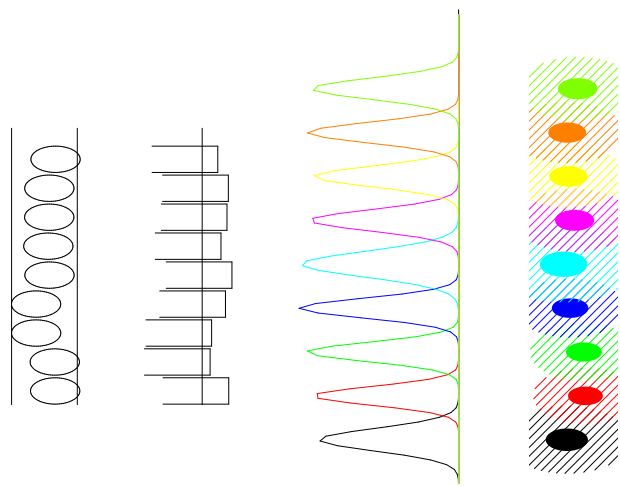}}
\caption{The left panel shows two sketches of the distribution of the fibers
  in the plane of the pseudo-slit, as viewed from both the front and the side.
  Both projections show the misalignments of the fibers. The right panel
  shows a scheme of the spectra projected in the CCD plane. It presents an
  intensity cut along the cross-dispersion. The intensity profile of each
  spectrum is shown a different color, illustrating the cross-talk
  effect.  The spots simulate emission lines as they are seen in the raw data
  (see \ref{fig:1}), with 1$\sigma$ cut represented as solid and 3$\sigma$ as
  hashed ellipses. They illustrate the effects of the misalignment on the
  dispersion.}
\label{fig:2}       
\end{figure}

\section{The raw data of fiber-fed spectrographs}
\label{sec:raw}

In most cases, the raw data of fiber-fed spectrographs consist of a collection
of spectra distributed along a certain axis of a 2D frame.
Figure \ref{fig:1} shows an example of IFS raw data, corresponding to PMAS
(\cite{roth05}) in the PPAK mode (\cite{kelz06}), illustrating this
distribution. Each spectrum is distributed in one direction with wavelength,
along the so-called dispersion axis. For each wavelength, it is also spread
along the perpendicular or ``cross-dispersion'' (or ``spatial'') axis,
following a characteristic profile of finite width, which may be considered
Gaussian in the simplest approximation.
If the spectra are tightly packed, contamination occurs among neighbors.  This
is referred to as cross-talk (see Fig. \ref{fig:1}).  Due to the mapping of
the fibers, adjacent spectra at the CCD may originate from distant locations
in the sky plane.  The cross-dispersion profile varies with position along the
pseudo-slit, generally being more symmetrical in the center.  The profile also
varies along the dispersion axis (e.g., \cite{beck00}; \cite{beck01}).
Spectrographs are generally designed to minimize the effects of cross-talk
(e.g., \cite{bers04}).  However, there is always a trade-off between the
number of fibers ``packed'' in the spectrograph, the size of the
CCD and the width of the profiles, and the final cross-talk. In addition, the spectra
are never perfectly aligned along the dispersion axis due to the configuration
of the instrument, its setup, the instrument focus, and any flexure effects.
The transmission also changes fiber-to-fiber due to 
different tensions, slight misalignments, and intrinsic physical differences. 
These imperfections also affect the width of the projected profiles along both the
cross-dispersion and the dispersion axes, and introduce small differences in
the spectral dispersion from fiber to fiber.

Figure \ref{fig:2} illustrates all these effects.  The left panel shows a
sketch of the distribution of the fibers in the plane of the pseudo-slit,
as viewed from both the front and the side. Due to the misalignments of the
fibers with respect to the axis of the pseudo slit, the light coming from each
fiber enters the spectrograph with small offsets relative to one another.  Therefore,
the dispersion solution varies slightly from fiber to fiber.
Meanwhile, the offsets with respect to the focal plane produce differences
in the profiles of the projected spectra at the CCD, as described before.  The
right panel shows these projections in the CCD plane.  It presents
an intensity cut along the cross-dispersion, where the intensity profile of
each spectrum is shown a different color. It illustrates the cross-talk,
the variations in focus, and the fiber-to-fiber transmission.  The spots
simulate an emission line as seen in the raw data, illustrating the
differences in the dispersion solution.  The same effect can be seen
directly in the raw data (Fig. \ref{fig:1}).

In any spectrograph, the spectral resolution changes from the center of the
slit to its edges. This is due to optical aberrations, which vary as a function of
field angle and of wavelength. This well-known effect is further complicated
by the differences among the fibers, as discussed above. 
The process of sky subtraction is particularly problematic, not only
because adjacent spectra at the CCD may correspond to distant locations on the sky, 
but also due to the above-mentioned differences in the dispersion
solution from fiber to fiber. All of these instrumental effects have to be
addressed in the data reduction.

\section{Procedures for the data reduction}
\label{sec:red}

All fiber-fed spectrographs yield similar raw data.
Thus the data reduction process generally follows the same sequence of steps,
regardless of the specific instrument involved.
As a starting point for the reduction,
it is assumed that the data have been bias and flat-field corrected
(when possible), with an external tool.
After these corrections, the data reduction consists of: (a) identification of
the position of the spectra on the detector for each pixel along the
dispersion axis, (b) scattered light subtraction and extraction of each
individual spectrum, (c) distortion correction of the extracted spectra, and
determination of the wavelength solution (i.e., dispersion correction), (d)
correction of the differences in the fiber-to-fiber transmission, (e) flux
calibration, and (f) sky emission subtraction. For IFUs it is also required to
(g) re-order the spectra on their original location in the sky and (h) correct
for the differential atmospheric refraction (DAR). Each of these reduction
steps has been treated with a set of command-line routines in {\tt R3D}. For
clarity, we will assume hereafter that the dispersion axis is the X-axis and
the cross-dispersion the Y-axis (although this is not a limitation of {\tt
  R3D}).

\subsection{Finding \& tracing the position of the spectra on the raw frame}

The location of the spectra is found at a certain column on the CCD by
comparing the intensity at each row along the column (or a coadded set
of columns around it) with that of $n$ adjacent pixels, looking for
those pixels that verify the maximum criteria, while requiring a minimum
distance from the previous maximum. Let be $i$
the pixel in the dispersion axis indicating the selected column in the
CCD, the $j$-pixel is the location of the peak number $k$ if it is the
$k$th pixel that verifies:

\begin{tabular}{l}
$I(i,j-n)<...<I(i,j-1)<I(i,j)$ and\\
$I(i,j)>...>I(i,j+n-1)>I(i,j+n)$\\
\end{tabular}

and 

$|j_k-j_{k-1}|>min\_dist$

 where $min\_dist$ is the minimum distance
between adjacent maxima, and $I(i,j)$ is the intensity in the raw data
at the $(i,j)$ location.

Once the location of the peak intensity is determined at the
resolution of one pixel, a more accurate centroid of that peak
($y_{peak}$) is determined with a simple parabolic maximum
determination.

 $y_{peak}=(j+1)-([I(i,j+1)-I(i,j)]/d+0.5)$

where 

 $d=[I(i,j+1)-2 I(i,j)+I(i,j-1)]$
 
 The result is stored as an ASCII file, containing an identification number
 for each fiber/spectrum and the pixel and centroid position of the peak
 intensity. The process is cross-checked visually by a graphical output. Lost
 spectra, due to broken fibers or low contrast between adjacent pixels, or
 misidentifications due to cosmic rays can be handled by editing this ASCII
 file. It is also possible to refine the input parameters based on the visual
 inspection of the output to remove misclassified identifications and to
 increase the contrast. The input parameters to be modified are the minimum
 distance between adjacent pixels, the number of adjacent pixels to
 check for the maximum criteria, the number of columns in the dispersion axis
 to coadd to increase the signal-to-noise/contrast and the initial column to
 look for the peaks. E.g., cosmic-rays can be identified as peaks of a
 particular spectrum when the number of adjacent pixels used to identify the
 location of the peaks ($n$) is too low. Two possible solutions to this
 problem could be (i) to enlarge this number, forcing the procedure to
 identify not just a local maximum, but a profile, or (ii) to increase the
 width of coadded columns to blur the cosmic-rays.

The adopted procedure to identify the location of the spectra projected at a
certain column of the CCD is independent of the pattern of the spectra
spacing, i.e., if it is uniform/regular or not or if there are gaps in between
different fibers. Indeed, it is not required to pre-define a certain pattern.
However, a certain knowledge of this pattern (or the disposition of the fibers 
in the pseudo-slit) is required to visually discriminate between real
spectra and misclassifications. 
Alternatively, one may use an external tool to create the ASCII file 
containing the location of the spectra on the CCD at each column.
This may then be cross-correlated with the real data 
and/or used as the input for the tracing routine. 
This option may be particularly appealing for instruments with many broken 
or low-transmission fibers like VIMOS.
 
Next, the tracing of the peak intensity along the dispersion axis is
performed by looking for maxima around each original location within a certain
window. The process is iterative, starting from the original column
and continuing to the end of the CCD.  If no maximum
is found within the predefined window for a certain column, the location of
the spectra on the previous column is used assuming a smooth behavior of the
trace along the dispersion axis. This method solves the possible
problems of bad columns, low sensitivity pixels, and/or cosmic rays.  Again,
the user may adjust the width of the input window, the number of adjacent
pixels used to look for the maxima criteria, and the number of coadded columns to
fine-tune the tracing procedure.  The result is stored as a 2D image, where
the X-axis corresponds to the original dispersion axis and the Y-axis
indicates the traced spectrum number. The stored value at each pixel is the
location of the peak centroid in the original frame of the given spectrum for
each pixel in the spectral direction.

The tracing procedure must be applied over well illuminated continuum
exposures. In many cases the science frames do not have enough signal-to-noise
through all the fibers at any wavelength for an accurate determination of the
tracing (e.g., emission-line dominated sources), and dome-flat exposures must
be used (or similar ones). In the case of IFUs that suffer from flexures,
these exposures may be taken inmediately before and after the science
exposures, and without moving the telescope from the position where the
observations have been performed.

We tested the peak identification and tracing routines with simple simulated
data, assuming Gaussian profiles with different FWHMs for the projection of the
spectra along the cross-dispersion axis.
The peak-to-peak separation was held constant.
The location of the spectra at the CCD is recovered within 1 pixel whenever
there is enough signal-to-noise per pixel (i.e., contrast between adjacent
peaks). There is no typical value for the signal-to-noise to be considered
good for the tracing procedure, since this process depends on other
parameters, like the FWHM of the profiles along the cross-dispersion, the
distance between adjacent spectra, the number of pixels compared for
the maximum criteria, and the number of coadded columns. 

\begin{figure*}
\centering
   \centering
\resizebox{\hsize}{!}
{\includegraphics[width=\hsize,angle=-90]{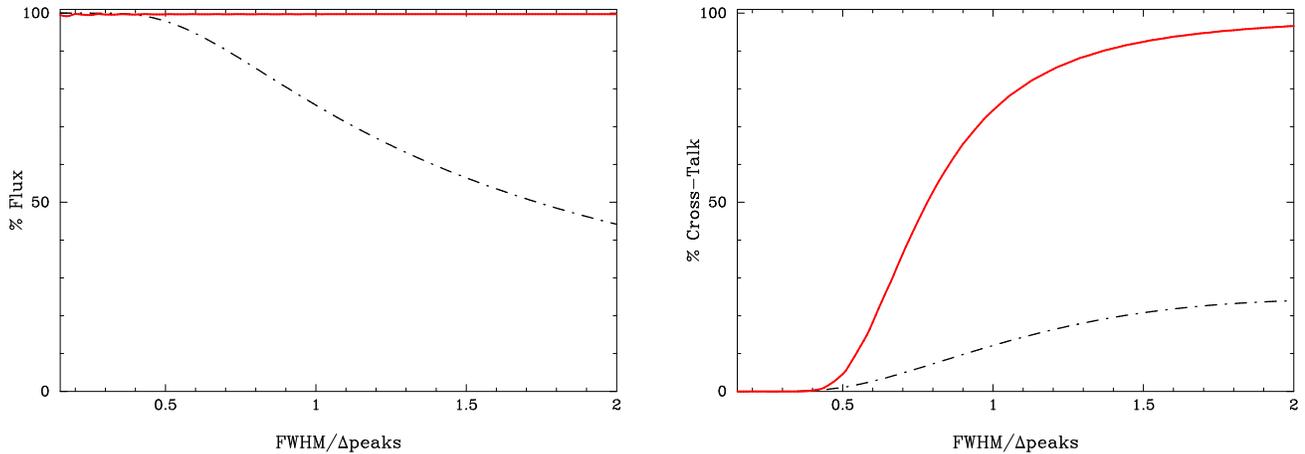}}
\caption{Results from the simulations. Left-panel: Fraction of the input flux
  recovered by an aperture extraction of a set of simulated spectra
  plotted versus the ratio between the FWHM of the profiles along the cross-dispersion
  axis and the distance between the peak intensity of adjacent spectra
  ($\Delta peaks$). The two lines shows the results for two different
  apertures. The dot-dashed line shows the results for a fixed aperture of the
  distance between adjacent spectra. The solid-line shows the results for a
  variable aperture of 3 times the FWHM.  Right-panel: Fraction of the input
  flux assigned by an aperture extraction to the adjacent one (i.e., the
  cross-talk) for the two different adopted apertures described above.}
\label{fig:3}       
\end{figure*}


\subsection{Scattered light subtraction}

Many spectrographs suffer from scattered light that contaminates the
science and calibration exposures. This light comes from reflections in
the instrument or light entering but not following the defined
light-path. It is normally identified as a smooth background
in-between the projected spectra on the raw data. It normally depends
on the input illumination, and it has a greater effect on twilight sky
exposures or long exposure time science frames in bright nights. 
To correct for this, information between the projected spectra
is required.  Pixels affected by the projected spectra are masked out 
and the remaining background is interpolated to
recover a map of the scattered light for the full frame. In {\tt R3D}
a scattered light map is obtained using the result from the tracing
algorithm, masking the pixels of the input image within a defined
aperture around the location of the spectra peaks for each pixel in
the dispersion axis. Once masked, the remaining pixels are used to
perform a polynomial function fitting along the cross-dispersion
axis. The derived frame is then smoothed to estimate the scattered
light map and subtracted from the input raw data.

The accuracy in the determination of the scattered light depends
strongly on the fiber-packing, i.e., the way the fibers are ordered in
the pseudo-slit. It is well determined if there is enough space in
between fibers and/or the fibers are packed in groups with well defined
separations. If the fibers are crowded, without {\it clean} spaces in
between, and suffering from strong cross-talk contamination, it is
almost unfeasible to estimate the scattered light.

\subsection{Spectra extraction}

After tracing the location of the spectra at the CCD, and subtracting the
scattered-light (if needed), the next reduction step is to extract the flux
corresponding to the different spectra at each pixel along the dispersion
axis. The most simple method to perform this extraction is to coadd the flux
within a certain aperture around the location of the spectra peaks in the raw
data, using the tracing information, and storing the resulting spectra in a 2D
image. This procedure is called aperture extraction, and it has been
implemented in {\tt R3D}. The X-axis of the resulting image corresponds to the
original dispersion axis, while the Y-axis corresponds to the ordering of the
spectra along the pseudo-slit. This is the so-called row-stacked spectra
representation (RSS, \cite{sanc04}). However, in most cases aperture
extraction is not the optimal method to recover the flux corresponding to each
spectrum. The distribution of the flux following pseudo-Gaussian profiles for
each spectrum along the cross-dispersion axis limits the accuracy of the 
recovered spectra.

Figure \ref{fig:3}, left panel, shows the percentage of flux recovered with an
aperture extraction for different FWHMs of the profiles along the
cross-dispersion axis (assuming Gaussian profiles) in units of the distance
between adjacent spectra ($\Delta$peaks, also known as {\it spectral pitch}).
The black dot-dashed line corresponds to a fixed aperture of the width of that
distance, while the red solid-line corresponds to a growing aperture of 3
times the FWHM. In the first case there is an underestimation of the recovered
flux, that grows with the FWHM of the profiles. This underestimation may be
solved by enlarging the aperture, as shown in the second case. However, the
contamination from flux coming from adjacent spectra (i.e., cross-talk) limits
the size of the selected aperture.  Figure \ref{fig:3}, right panel, shows the
percentage of flux of a given spectrum that is erroneously assigned to the 
nearest adjacent spectrum by an aperture extraction.  This
percentage is plotted as a function of the FWHM of the instrumental profiles,
where the two lines again correspond to the apertures described above.
The cross-talk percentage grows with the FWHM for both aperture selections. It
is important to note here that cross-talk is an incoherent contamination, and
in many science cases and/or instrument designs it is preferable to keep it as
low as possible: an average value of $\sim$1\% with a maximum of a $\sim$10\%
(e.g., \cite{bers04}). For an aperture of $\sim$$\Delta$peaks, this level of
cross-talk is found for FWHMs$\sim$$\Delta$peaks.  However, in the case of an
aperture that maximizes the recovered flux, the level of cross-talk is already
$\sim$60\% for the same ratio between FWHMs and $\Delta$peaks.

The need for packing the largest possible number of spectra on CCDs
with limited sizes ensures a certain level of cross talk.
The FWHM of the projected profile along the
cross-dispersion axis is normally defined by the design of the spectrograph
and the size of the input fibers. Therefore, the ability to pack
spectra with a reduced cross-talk is controlled by the ratio between
these FWHMs and the defined $\Delta$peaks. A ratio of
FWHM$\sim$0.5$\Delta$peaks seems to be a good compromise for
maximizing the packing of spectra at the CCD, without large cross-talk
contamination. For illustration, PMAS/PPAK spectra profiles have a
FWHM of the order of $\sim$2.3 pixels, and a $\Delta$peaks of $\sim$5
pixels, in the 2$\times$2 binning mode (\cite{kelz06}).  Selecting an
aperture size of the order of $\Delta$peaks seems to be an
acceptable compromise between maximizing the recovered flux and
minimizing the cross-talk, for aperture extractions.  However, in some
cases, when the raw frames are too crowded with spectra and the
cross-talk is severe, this approach is not valid (e.g., \cite{murr00}).
On the other hand, there are spectrographs for which the $\Delta$peaks
is much larger than the FWHMs (e.g., PMAS in Larr mode, without
nod\&shuffle, \cite{roth05}). In this case it is possible to maximize the
fraction of recovered flux by enlarging the aperture, with a reduced
cross-talk contamination.

\begin{figure*}
\centering
   \centering
\resizebox{\hsize}{!}
{
\includegraphics[width=\hsize]{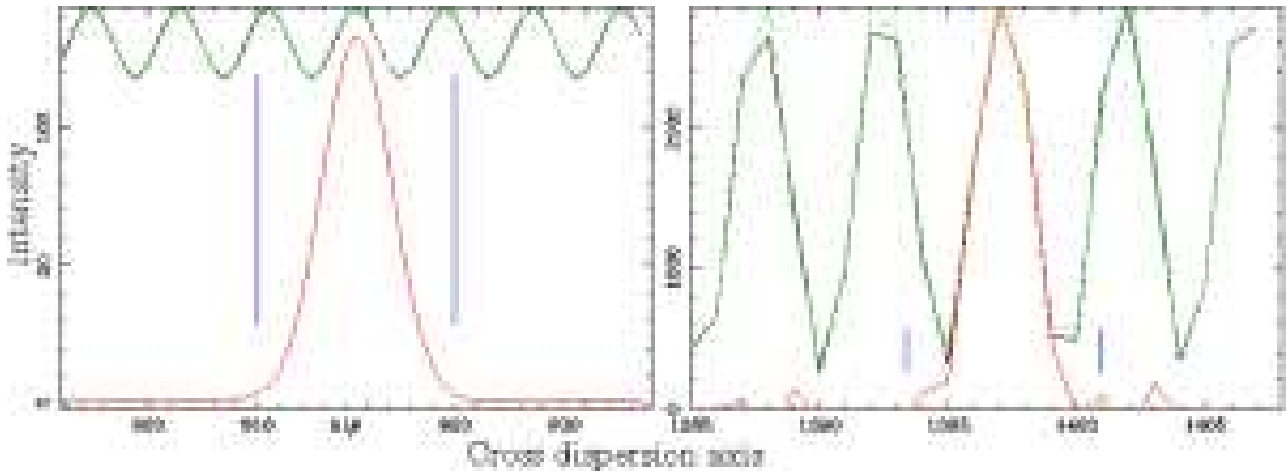}
}
\caption{Left-panel: Section of an intensity cut along the cross-dispersion of a simulated
  set of fiber-fed spectra (black solid-line), created assuming a Gaussian
  profile with a FWHM equal to the distance between adjacent fibers, i.e., a
  system with a strong cross-talk. No noise has been included in the
  simulation. The green dot-dashed line shows the modeled profiles created by
  the {\it Gaussian-suppression} technique, and the red solid-line indicates
  the recovered profile of the considered spectrum once decontaminated from
  the contribution of the adjacent spectra by this technique, assuming an
  initial aperture indicated by the vertical lines. Right-panel: Similar plot
  for real data of a continuum illuminated exposure obtained with the PPAK
  mode of the PMAS spectrograph, a system with a FWHM$\sim$0.5 times the distance
  between adjacent fibers.}
\label{fig:4}       
\end{figure*}

\begin{figure}
\centering
   \centering
\resizebox{\hsize}{!}
{\includegraphics[width=\hsize]{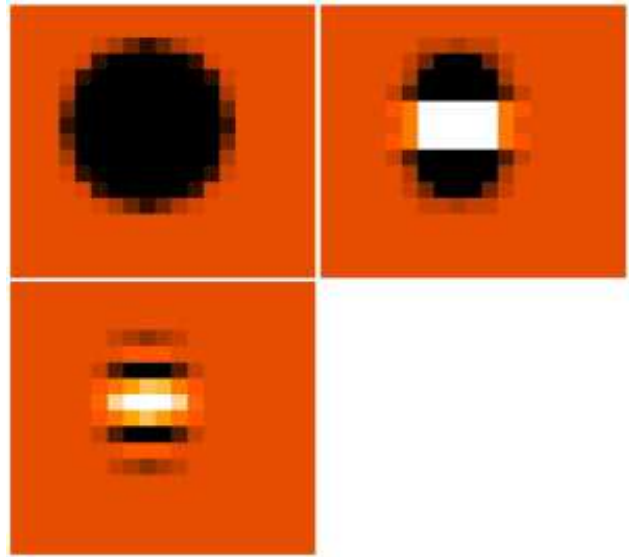}}
\caption{Top-left: 20$\times$20 spaxels intensity map of a simulated star
  with a Gaussian profile of 3 spaxels FWHM. Top-right: Result of the
  subtraction of the former map from the reconstructed map created after
  simulating a set of IFS data with the previous intensity distribution and
  {\it reducing} them using a pure aperture extraction. Bottom-left: Similar
  map obtained {\it reducing} the simulated data with a {\it
    Gaussian-suppression} extraction. The same grayscale range and scale
  intervals were used in the two latter panels.}
\label{fig:cross}       
\end{figure}

The cross-talk may have strong implications in the interpretation of the data.
In a general case, the adjacent spectra on the plane of the CCD may not
correspond to nearby locations in the sky plane (e.g., PMAS and VIMOS). On the
other hand, other instruments, like PMAS/PPAK and INTEGRAL (\cite{ar98}), have
calibration-fibers located in between the science ones along the pseudo-slit.
In both cases the cross-talk mix-up spectra from locations that may be
not physically related or from spectra with a completely different nature.

Different techniques have been implemented to minimize the cross-talk. The
full version of {\tt P3d} (\cite{beck01}, \cite{roth05}) uses an instrumental
profile database to decouple the flux coming from adjacent spectra. This
method is rather slow, and it has the disadvantage of requiring an accurate
determination of the profiles. This is not always possible due to instrumental
designs, and they may even change during the observing run. A more simple
approach is to assume a Gaussian profile and perform a multi-Gaussian fit. We
implemented this method in {\tt R3D}, but it turned out to be too slow, and
strongly dependent on the signal-to-noise of the data.

We developed a new technique that reduces the effects of the cross-talk and
maximizes the recovered flux, named {\it Gaussian-suppression}. The technique
works assuming that the spectral profiles are well approximated by Gaussian
functions, and that the FWHMs of all of them are almost equal (this last
assumption is not a limitation of the method, but a simplification). In a
first iteration it performs an aperture extraction, as described above. 
Then for each spectrum the flux corresponding to $n$ adjacent spectra is
subtracted assuming a Gaussian function with the defined FWHM, and with the
integrated intensity derived by previous aperture extraction (a multi-Gaussian
approximation). Once the flux from adjacent spectra is subtracted, a new
aperture extraction is performed with the same or a different aperture.

Figure \ref{fig:4} illustrates the method.  In the left panel, the solid black line
shows a cut along the cross-dispersion axis for a simulated IFS raw frame
created assuming Gaussian profiles for the projection of the spectra on the
CCD, with a FWHM equal to the size of $\Delta$peaks (i.e., data with strong
cross-talk). The green dashed-line shows the multi-Gaussian approximation,
whose intensities are derived from the aperture extraction performed in the
first iteration. The red solid-line shows the recovered profile after
suppressing the flux coming from adjacent spectra. The method works well for
severe spectral packing in the CCD. Figure \ref{fig:4}, right panel, shows a
similar plot for real data obtained with PMAS/PPAK. It illustrates the
performance of the method in the presence of noise and for profiles without
Gaussian shapes.

\begin{figure*}
\centering
   \centering
\resizebox{\hsize}{!}
{\includegraphics[width=\hsize]{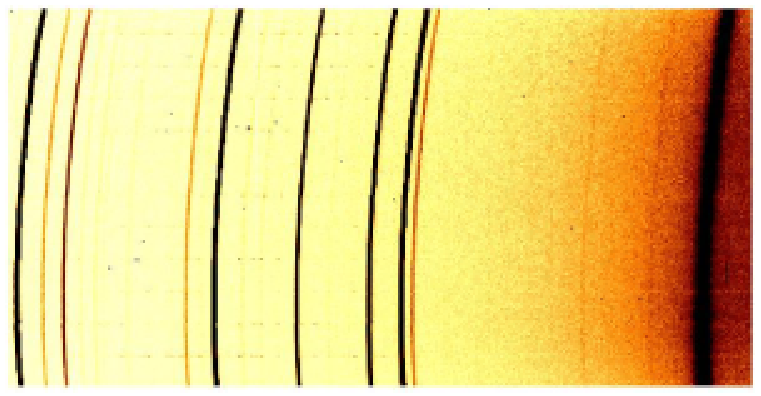}
\includegraphics[width=\hsize]{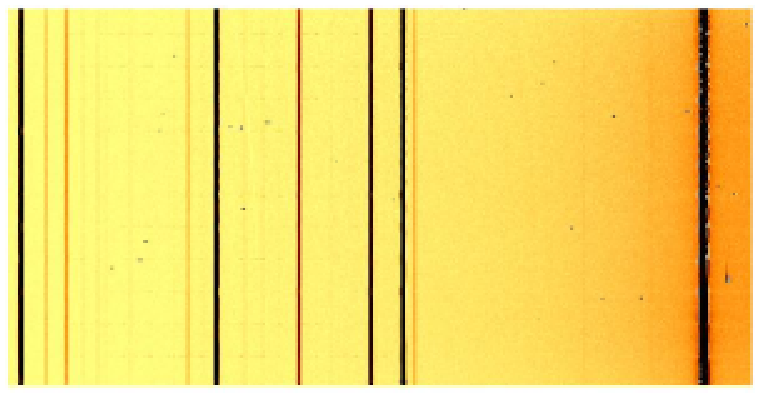}
}
\caption{Left panel: Example of extracted spectra of a calibration arc lamp
obtained with PMAS in the PPAK mode. Right panel: Same data after wavelength calibration.}
\label{fig:5}       
\end{figure*}

We strongly recommend the user to test the {\it Gaussian-suppression}
technique with simulations that resemble as much as possible the science data
before using it. The required input parameters are strongly dependent on: (1)
the shape of the spectra projected along the cross-dispersion, (2) the
distribution of the spectra at the CCD (e.g., homogeneously packed, packed in
groups, or more or less randomly distributed), and (3) the position table of
the fibers in the sky plane (i.e., if adjacent spectra at the CCD corresponds
to adjacent positions in the sky).  As a probe we simulated IFS data of a
fiber-fed IFU with a lensarray of 20$\times$20 spaxels, distributed in a
regular grid with a square pattern (Fig \ref{fig:cross}). The spectra are
projected along the cross-dispersion axis following Gaussian profiles of
FWHM=3.5 pixels, with a distance of 7 pixels between the peak emission of
adjacent spectra. The simulated data is comprised of a model star with a
Gaussian profile in the spatial direction with a FWHM of 3 spaxels.  No noise
was included in this simulation. The simulated spectra are {\it reduced} using
{\tt R3D}, following the steps described above. The extraction was performed
twice, using both a pure aperture extraction and the {\it
  Gaussian-suppression} technique. In the first case we used a 7 pixel
aperture, following our previous results (Fig.  \ref{fig:3}). The flux is
recovered within $\sim$10\% of the original values for any spaxel at any
wavelength using an aperture extraction, and with a standard deviation of the
residuals of $\sim$2\% of the original flux. In the second case we tested
different combinations of parameters to recover as closely as possible the
original integrated flux. We finally use an initial aperture of 6 pixels and a
final one of 9 pixels. The original flux is recovered within $\sim$5\%, with a
standard deviation of the residuals of $\sim$0.5\%.

Cross-talk between spectra, in combination with the mapping scheme of
the spatial elements, causes distortions in the re-constructed images.
In our simulated data the position table is sequential, from left-top
to right-bottom, following the rectangular grid. In instruments with
this kind of position table the re-constructed images are elongated
(e.g., \cite{roth04}). Figure \ref{fig:cross} illustrates this
effect. The top-left panel shows the input intensity map of the
simulated star. The top-right panel shows the subtraction of this
original map from the reconstructed map of the {\it reduced} data
using a pure aperture extraction. As expected there is a transfer of
flux from the peak emission along the vertical axis, that produces a
flattening and elongation of the shape. The bottom-left panel shows a
similar map for data {\it reduced} using the {\it Gaussian-suppression}
extraction. The original shape is reconstructed better.

\subsection{Distortion correction and wavelength solution}

For grating spectrographs, the entrance slit is distorted and imaged
as a curve onto the CCD (e.g., \cite{mea84}).
When fed with fibers, additional distortions are added to this intrinsic
curvature due to the placing of the fibers in the pseudo-slit, as described in
Section \ref{sec:raw}, Fig. 2. 
Figure \ref{fig:5}, left panel, shows an example of extracted spectra where
the curvature and distortions of the dispersion solution are seen.  The data
correspond to a calibration arc-lamp observed with PMAS in the PPAK mode.
The discontinuities and slight differences in the dispersion solution
fiber-to-fiber prevent us from performing 2D modeling of the distortion map
with an analytical function, nor at the level the extracted spectra,
nor at that of the raw data.  These {\it distortions} must be
corrected fiber-to-fiber before finding a common wavelength solution.

{\tt R3D} performs this correction in a two-step procedure, using arc
calibration lamp exposures (like the one in Fig.\ref{fig:5}). In the first step,
the peak intensity of a single emission line is traced along the
cross-dispersion axis, and shifted to a common reference, by a linear shift.
After that, the intensity peak of a set of selected emission lines is traced,
and a polynomial distortion correction is determined to recenter all the
lines to a common reference. The distortion correction maps are stored to be
applied subsequently to the science exposures. Figure \ref{fig:5}, right panel,
shows the same extracted spectra of the left panel after applying the
distortion correction. A one dimensional spline interpolation algorithm is
applied spectrum-to-spectrum in this process.

In the second step, the wavelength coordinate system is determined by identifying the
wavelengths of the arc emission lines, using an interactive routine. The
distortion-corrected spectra of the arcs are then transformed to a linear
wavelength coordinate system by a one dimensional spline interpolation,
assuming a polynomial transformation between both coordinate systems. The
required transformation is stored in an ASCII file to be applied over the
science data.

\begin{figure*}
\centering
   \centering
\resizebox{\hsize}{!}
{\includegraphics[width=\hsize]{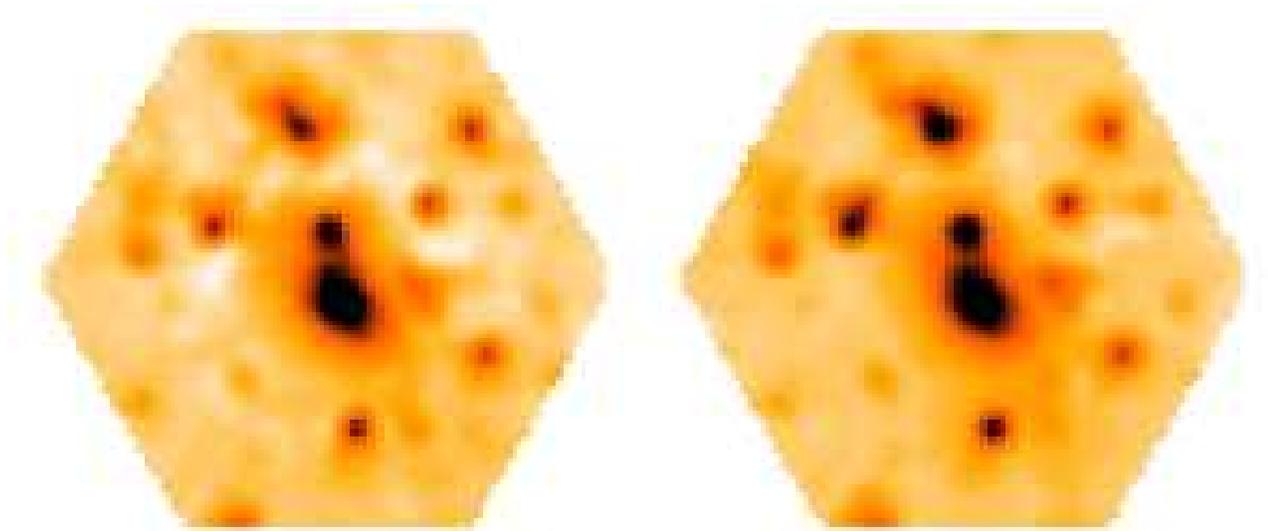}}
\caption{
  Example of the effects of flat-fielding. The panels shows a map, extracted
  from a reduced and interpolated dataset of PPAK.  Some structures seen in
  the left panel are artefacts that disappear after calibration of the
  differences in the fiber-to-fiber transmission.  }
\label{fig:flat}       
\end{figure*}


The accuracy of the final wavelength solution depends on the selected order of
the polynomial function, the number of identified lines, and the coverage of
emission lines along the wavelength range. These values depend strongly on the
instrument and setup. For example, for PMAS in the PPAK mode, using the lowest
resolution (e.g., V300 grating, \cite{roth05}), it is possible to achieve an
accuracy of $\sim$0.15\AA\ ($rms$ of the fit) identifying just 9 emission
lines and using a polynomial function of order 4. A similar accuracy is
obtained for VIMOS in the High-resolution Orange mode by identifying 26
emission lines and using a polynomial function of order 5.

\subsection{Fiber-to-fiber response correction}

For most IFUs, the transmission varies appreciably fiber-to-fiber.
To correct for this, an exposure of a continuum, well-illuminated, 
and flat source is required, like a dome or sky flat.
This exposure is used to determine the differences in the
transmission fiber-to-fiber, by comparing the flux spectra-to-spectra,
and deriving the so-called fiber-flat. In certain IFUs the fiber-flat shows a
strong wavelength dependency, while in others this is unappreciable. {\tt R3D}
includes tools for deriving both kinds of fiber-flats.  In the first case the
spectra are divided by the median spectrum, obtaining a normalized fiber-flat,
including the wavelength dependence. In the second case, the flux within a
selected wavelength range is averaged for each spectrum and divided by the
median of these values for all the spectra. In both cases, the science spectra
must be divided by the obtained fiber-flat.

Figure \ref{fig:flat} illustrates the importance of the fiber-flat correction.
It shows a reconstructed map extracted from a datacube of PPAK data of the
core of the galaxy cluster Abell 2218 (\cite{sanc06}), centered at
$\sim$5780\AA. The spectra were reduced with {\tt R3D}, and rearranged in
their spatial distribution and interpolated to create a regular grid datacube
using {\tt E3D} (\cite{sanc04}), as discussed below.  Left and right panels
show the map before and after correcting for the differences in the
fiber-to-fiber transmission. Note that before the correction, the image
has an artificial ring-like structure, which is almost completely removed in the right
panel. This is obviously seeing in IFS reconstructed maps, but it is difficult to
identify in MOS spectroscopy. In both cases, the result cannot be trusted
before correcting for the differences in the fiber-to-fiber response.  
This affects not only the comparison between adjacent spectra and the
reconstructed 2D flux distributions, 
but it also strongly affects the sky-subtraction.

\subsection{Flux calibration}

Once the spectra are extracted, corrected for distortions, wavelength
calibrated, and corrected for differences in transmission fiber-to-fiber,
they must finally be flux calibrated. This calibration requires the observation
of a spectrophotometric standard star during the night. As in slit-spectroscopy,
where slit losses impose a severe limitation, absolute spectrophotometry with
fiber-fed spectrographs is complex. Fiber-fed spectrographs suffer light
losses when the fibers are smaller than the seeing-disc (or the astronomical
target) and the calibration standard stars are not completely well centered in
a single fiber. Both problems are better handled by IFUs, where the
astronomical targets and the calibration stars are sampled by different
fibers. However, IFUs based on pure fiber-bundles, like the PPAK mode of PMAS
or INTEGRAL, do not cover the entire field-of-view (instead having typical
fill-factors of $\sim$60-70\%), which imposes flux losses. Other IFUs, like the
PMAS/Larr or VIMOS, have solved this problem coupling a lensarray to the
fiber-bundle. However, in most cases, it is not possible to obtain in a given night
all the observations of spectrophotometric standards required to
perform reliable absolute spectrophotometry. The most extended approach to
solve the problem is to determine a relative spectrophotometry, and
recalibrate the spectra later, using additional information, like broad-band
photometry (e.g., \cite{bego05}, \cite{sanc06b}).

{\tt R3D} includes a tool that compares the measured spectrum of a 
spectrophotometric standard star with the absolute values, convolving
the spectra with a Gaussian kernel to degradate and match their
resolutions. The ratio between both spectra at each wavelength, 
or the instrumental response, is
stored and subsequently applied to the science frames to
flux-calibrate them.

\subsection{Sky Subtraction}

The night sky emission spectrum must be subtracted from the science spectra.
This subtraction should be performed before the flux calibration, if an
external sky-frame is used, or can be done after it, if the sky is derived
from the same dataset.  In long-slit spectroscopy the sky is sampled in
different regions of the slit (normally along the edges).  In the most general
case a median sky spectrum is obtained by averaging the spectra of these
regions or interpolating (or extrapolating) them to the regions of the slit
that sample the astronomical target. This is possible due to the size of the
long-slits (typically several arcmins), compared with the size of the
astronomical objects of interest.  Unfortunately, the field-of-view of
currently available IFUs is much smaller, being of the order of a few arcsecs
to one arcmin (in the case of VIMOS or PPAK).  If the astronomical objects
fill all of the field-of-view, no spaxels sample any blank sky. In this case
it is required to obtain supplementary sky exposures, and perform a direct
subtraction of the reduced frames. On the other hand, if not all the
field-of-view is filled with objects, it is possible to select those spectra
free of contamination from objects, average them, and subtract it from the
science spectra.  This is simple to do using additional tools, such as {\tt
  E3D} (\cite{sanc04}).

\begin{figure*}
\centering
   \centering
\resizebox{\hsize}{!}
{\includegraphics[width=\hsize]{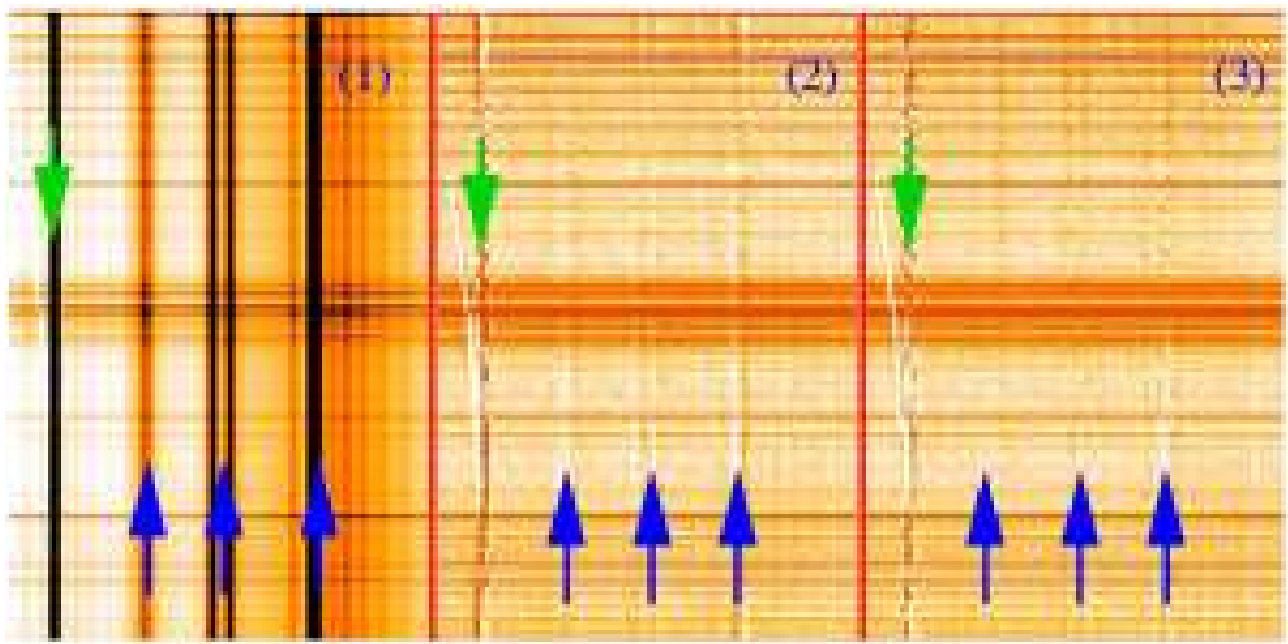}}
\caption{
Example of different methods for the sky subtraction. The left panel (1) shows reduced row-stacked spectra obtained with PMAS/PPAK. The central
panel (2) 
shows the same exposure after subtraction of an average sky spectra. 
For the right panel (3) the sky
spectrum was determined for each fiber by interpolating the sky
spectra to  
the location of the science ones. Both images in panels (2) and (3) are
plotted using the same grayscale. Arrows
  located at the top of the image indicate the sky-lines where the residuals
  are reduced, but still strong. Arrows located at the bottom of the image indicate the sky-lines were the residuals are clearly decreased.}
\label{fig:sky}       
\end{figure*}

Some spectrographs (e.g., PPAK, INTEGRAL) have designs that include additional
fibers that probe the sky far enough from the science field-of-view to avoid
contamination by the astronomical object.  These {\it sky} fibers are normally
distributed in between the science fibers across the entire pseudo-slit, in
order to avoid any bias caused by the spectrograph off-axis position.  {\tt
  R3D} includes routines to extract and average the spectra obtained through
these sky fibers, and subtract it from the science spectra.

As explained in section \ref{sec:raw}, imaging effects within the spectrograph
and offsets at the fiber-slit change the effective dispersion, and have to be
corrected by shifting, stretching, and interpolating the data to a linear
wavelength coordinate system. However,
this procedure does not correct for the differences in the shape of the
emission lines along the cross-dispersion axis. This effect is also present in slit
spectroscopy, and as a result, the sky-subtracted spectra show
characteristic residuals in the location of the sky emission lines. The effect
is more severe in fiber-based spectrographs, since the effective dispersion
(and the shape of the emission lines) varies fiber-to-fiber. Therefore,
even in the case of IFUs with {\it sky} fibers, an accurate decontamination
from the sky emission is complex. {\tt R3D} includes a specific tool for IFUs
with {\it sky} fibers that determines the sky spectra corresponding to any
science fiber by an interpolation of the spectra obtained through the {\it
  sky} ones.  This tool increases the accuracy of the sky subtraction,
compared with the results obtained by a simple averaging of the {\it sky}
spectra.  It takes into account the distortion in the wavelength solution
along the cross-dispersion. However, it is still not optimal since it does not
take into account the individual differences in the dispersion fiber-to-fiber.
Figure \ref{fig:sky} shows an example of both sky subtraction procedures. More
sophisticated procedures to subtract the sky have been performed in the data
reduction techniques of different instruments (e.g., \cite{bers05}),
but these are not currently implemented in {\tt R3D}.

A more precise sky subtraction method is the
nod \& shuffle mode (e.g., \cite{roth05}, \cite{alli02}), 
that almost simultaneously records both sky and object spectra, 
using the identical optical path for both. 
While this method increases the accuracy of the sky subtraction, it 
doubles both the required exposure time, and the number of spectra at 
the CCD. {\tt R3D} is also capable of reducing this type of data.

\subsection{Re-ordering the spectra on the sky}
\label{sec:map}

As explained above, fiber-fed spectrographs and IFUs conduct light to the CCD
from different positions in the sky.
In both cases, but for IFUs in particular,
it is important to re-order the spectra to their original positions in
the sky. 
A position table is required, relating the spectra to their locations in the sky.
Given this additional information, it is possible to use specific tools for IFS,
like E3D (\cite{sanc04}), to reorder the spectra. Then, it is possible
to interpolate the data spatially and create a regular gridded
datacube, and reconstruct the original image of the target at any
wavelength, like the one shown in Fig. \ref{fig:flat}. Spatial
interpolations are required in many cases, for example, to correct for
differential atmospheric refraction (Sec. \ref{sec:dar}). This
solution has been widely used in IFS, to become almost a standard
procedure.

Some IFUs have lensarrays coupled to the fiber-bundles to avoid gaps
in the coverage of the field-of-view (e.g., PMAS, VIMOS, GMOS). In
some of them the lensarray already has a regular grid rectangular
pattern (PMAS, VIMOS), while others have bee-cell-like hexagonal
patterns (GMOS). In both cases spatial interpolation is not recommended
in reordering the spectra to their locations in the sky. We
included different procedures in {\tt R3D} to reorder 2D RSS
spectra of this kind of IFU into a datacube, following the pattern of
their position table. The adopted solution is trivial for IFUs with
regular grid rectangular lensarrays. This is not the case for
lensarrays with hexagonal patterns.

The adopted solution in {\tt R3D} for this latter kind of IFU is to create a
regular grid rectangular datacube with the desired spatial pixel and covering
the same wavelength range as the original 2D RSS frame. For each spatial
pixel at each wavelength the recovered intensity would be the original one, if
the pixel is included inside an hexagon cell, or an average of the intensities
of the covered hexagons, if the pixel is in between different hexagon cells
(weighted by the fraction of pixel area in each hexagon). Finally the obtained
datacube must be corrected for differences in the input and output apertures
(i.e., the ratio between the areas of the input and output spaxels). This
solution, that we call {\it regularization}, does not impose any
interpolation.

We used this {\it regularization} scheme to combine exposures taken
at different adjacent positions in the sky with the same IFU
(Mosaicing), with overlapping areas. It was also used to combine
dithered exposures, obtained to cover the entire field-of-view using
fiber-bundle IFUs (e.g., PPAK, \cite{sanc06}, \cite{sanc06b}). To our
knowledge, the previous solution generally adopted was to interpolate each of
the original pointings frames before combining. With that solution most of
the combined intensities come from interpolated values, not from measured
ones. With our adopted scheme, all the spectra result from the
combination of measured values. On the other hand, the depth of the
final datacube changes pixel-to-pixel in dithering schemes with low
numbers of individual pointings.

\subsection{Differential Atmospheric Refraction}
\label{sec:dar}

IFS is capable of performing a correction of the differential atmospheric refraction
(Filippenko 1982) after the observations, without requiring knowledge of the
original orientation of the instrument and without the need of a compensator
(\cite{medi}, \cite{roth04}). Reduced IFS data can be understood as a set of
narrow-band images with a band-width equal to the spectral resolution.  These
images can be recentered using the theoretical offset determined by the DAR
formulae (Filippenko 1982), or, as in 2D imaging, tracing the intensity peak
of a reference object in the field-of-view (or a DAR reference observation)
along the spectral range, and recentering it.  Note that this latter approach
is basically unfeasible in slit spectroscopy, this being one of the 
fundamental differences between the two methods.
The correction of the DAR is critical for the proper combination
of different IFS exposures of the same object taken at different altitudes and
under different atmospheric conditions. It is also fundamental for a proper
alignment of mosaic and dither exposures.

{\tt R3D} includes tools for tracing the location of the peak intensity of a
particular object in the field-of-view along the dispersion axis. These
locations are estimated by determining the centroid of the object in the image
slice extracted at each wavelength from the datacube. Then, it is possible to
shift the full datacube to a common reference by resampling and shifting each
image slice at each wavelength (using an interpolation scheme), and storing the
result in a new datacube.  It is important to note here that any DAR
correction imposes an interpolation in the spatial direction.

\section{Comparison with other tools}
\label{sec:com}

Most of the techniques discussed in this article are well known by
specialists, being implemented both in general packages like IRAF
(\cite{vald92,vald92b}), and in specific ones like PMAS (\cite{beck01,roth05}),
the VIMOS (\cite{vimos05,vimos05b}) and GMOS reduction package
(\cite{turn06}).  However, each implementation is somewhat different, 
affecting the reduction strategies and the results. It is
unfeasible to compare our adopted implementations with those of all the
existing packages.  Therefore, in order to illustrate them we compare briefly
{\tt R3D} and {\tt dofibers}, a generalist tool for reducing fiber-fed
spectroscopic data implemented in IRAF (\cite{vald92}).

{\tt dofibers}, like many other high-level tools in IRAF, compiles
in a single task a series of different individual ones that are called (or
not), in a sequential way. Each of these tasks corresponds, more or
less, to each of the reduction steps explained through this article.
As any other IRAF tool, it is possible to call it within a CL script, in a
sequential way, as a single task or recursively. In all these aspects
it seems remarkably similar to {\tt R3D}. However, it has considerable
differences.


As mentioned above, {\tt R3D} is comprised of a set of
subroutines, each one related to a particular step of the data reduction. It
is possible to modify any of the input parameters of each tool, storing the
intermediate results. Any particular task at any step of the data reduction
can be replaced with a similar one coded by the user and/or belonging to
another reduction package. The intermediate results, like locating the spectra
and tracing them, are amenable to simple modifications, making {\tt R3D} a
flexible tool. On the other hand, {\tt dofibers} is a monolithic task, which
does not allow full access to all the input parameters of the
lower-level tasks called during the reduction process.  Like any other IRAF
tool, it relies on entries in certain header keywords. To replace a single
task would require a deep knowledge of which headers modifies and the required
output values. Furthermore, the output of the intermediate reduction steps are
not stored in general and in many cases
are difficult or impossible to modify (like the output of the tracing procedure). 

Like in any other IRAF high-level tool, it is possible to use independently the
different low-level tasks called by {\tt dofibers}, creating your own
high-level tool useful for a particular instrument (e.g., {\tt doargus}, {\tt
  dofoe}, {\tt dohydra}...). These low level routines can be also modified,
simplifying their use. For example, {\tt int\_apall}, the main task of the INTEGRAL
reduction package (\cite{ar98}), is a simpler version of {\tt apall} (one of
the tasks called by {\tt dofibers}), with most of the required parameters
already pre-defined for each of the setups of this instrument.

\subsection{Comparison step by step}

Both {\tt R3D} and {\tt dofibers} require a pre-reduction of the raw data,
that may include bias and dark correction, overscan trimmering, and
pixel-to-pixel flat-field correction. In {\tt dofibers} this pre-reduction
must be done with another IRAF tool, {\tt ccdproc}. If not, the headers of the
input files must be modified.  In {\tt R3D}, this pre-reduction can be
performed with any external tool.


The {\tt dofibers} procedure to locate the pixel of the intensity peak of each 
spectrum along the cross-dispersion axis is similar to that of {\tt R3D}. However,
the exact procedure to determine the centroid of those peaks differs. {\tt
  dofibers} uses a sophisticated procedure included in {\tt center1d}, which
relies on the contrast between the peak intensity and the supposed background.
This procedure requires that the space in between fibers is large enough
compared to the FWHM of the pseudo-Gaussian profiles to allow enough contrast
of the peaks. As a result, it often fails to locate spectra in spectrographs
with heavy fiber packing and strong cross-talk (e.g., MPSF, \cite{beck01},
\cite{ingo05}). The solution adopted in {\tt R3D}, a parabolic maximum
determination, is less accurate, but more reliable in most cases.
In addition, the difficulty to predefine or modify the estimated location of the
spectra with {\tt dofibers} makes it unfeasible for use in the reduction of data from
spectrographs with many broken fibers and strong displacements of the location
of the spectra frame-to-frame (e.g., VIMOS).

The tracing of peaks along the dispersion axis is also different. In
{\tt dofibers} this is performed every fixed number of pixels, and not in all of
them (in general), as is done by {\tt R3D}. After this non-contiguous search, the
final trace is obtained by fitting the results with a polynomial function
(Legendre, Chebyshev, or spline). The combination of a non-contiguous
determination of the location of the spectra and the fitting of the results to
a polynomial function, may produce significant deviations from the real trace
if the input parameters are not handled with care.

The adopted solutions to determine and correct the scattered light
are similar in both packages, with the difference being that in {\tt
dofibers} a 2D fitting is performed over the unmasked pixels (pixels
not affected by signal from the spectra), while in {\tt R3D} we
perform a 1D fitting along the cross-dispersion axis for each pixel in
the dispersion axis. In both packages the final result is smoothed
before applying it.

The {\tt dofibers} spectra extraction can be performed, as in {\tt R3D}, 
by coadding the flux within a fixed aperture around the previously determined 
locations along the dispersion axis, also taking fractional pixels into account. 
{\tt dofibers} also includes other extraction methods, 
such as variance weight flux extraction 
and adaptive apertures based on the contrast of intensity of the
peaks (\cite{vald92b}). 
On the other hand, it does not include any procedure to handle the cross-talk.

Distortion correction and derivation of the wavelength solution are performed
in {\tt dofibers} using the IRAF tools {\tt identify}, {\tt reidentify} and
{\tt dispcor}. These tools allow use of (i) comparison arcs taken immediately
after each science frame, (ii) specific calibration spectra placed in between
the science ones (e.g., as in PPAK, \cite{kelz06}), and (iii) a series of arc
frames that can be interpolated. In case (i) the procedure is remarkably
similar to that adopted in {\tt R3D}. However, with our adopted solution it is
possible to use more emission lines than the identified ones to correct for
the distortions of the spectra and resample them to the same wavelength
solution. Case (ii) is strongly recommended not to be used, since the
wavelength solution must be derived on a spectrum-by-spectrum basis. If not,
it may produce substantial errors. Case (iii) may also be dangerous, since the
interpolation of different arcs is based on their distance in time with the
science frame (defined by the julian date). In spectrographs that show strong
flexures, like PMAS or VIMOS, this solution may produce strong deviations from
the appropriate distortion correction and wavelength solution.

Another difference in this step lies in the adopted functions included in
both packages to find the distortion correction and wavelength
solution. Both {\tt dofibers} and {\tt R3D} include spline and 
Chebyshev functions.  However, {\tt dofibers} does not include linear
polynomial functions, while {\tt R3D} lacks Legendre polynomials. The adopted
functions to interpolate and resample the spectra to a common linear
wavelength solution are also different. While {\tt R3D} only includes
spline interpolations, {\tt dofibers} includes three different
interpolation functions.

The adopted schemes for correcting the differences in fiber-to-fiber
transmission are similar in both packages, with the difference being in
{\tt R3D} it is possible to smooth the transmission ratio along the
dispersion axis prior to applying it. In most cases this ratio has a
smooth wavelength dependency, and a smoothing is recommended to
increase the signal-to-noise. This solution is also included in 
other reduction tools, such as {\tt P3d} (\cite{beck01}, \cite{roth05}).

The sky subtraction scheme adopted in {\tt dofibers} is the simplier
method included in {\tt R3D}, and consists of obtaining the median of
the spectra marked as sky to create a single sky spectrum that is then
subtracted from each spectrum in the dataset. This procedure, along with more
refined improvements, have been discussed in previous sections.
 
Finally, it is important to note here that {\tt dofibers} was designed to
handle fiber-fed spectroscopic data but not IFS data. Therefore, it
does not include any procedure to re-order the spectra in sky, correct
for DAR, or perform 3D dithering or mosaicing.


\section{Conclusions}
\label{sec:con}

This article has explained the particular characteristics of the raw
data and data reduction of fiber-fed spectrographs in general and IFUs in
particular. We presented a specialized package for reducing this kind of
data, {\tt R3D}, explaining in detail how the various steps in the data
reduction process have been implemented. The effect of cross-talk
and its possible solutions were discussed in detail, with simulations that
illustrate them. {\tt R3D} has been tested over several IFUs, including PMAS
in its Lensarray and PPAK modes (which differs substantially), GMOS, VIMOS and
INTEGRAL. Throughout this article we used PPAK data to illustrate how the data
{\it evolve} after passing through each reduction step. The basic adopted solutions for
the reduction steps were compared with those of the IRAF task
{\tt dofibers}, which is widely used to reduce this kind of data.

Integral Field Spectroscopy is evolving fast from a technique for
specialists to a common user technique, and generalist tools, like
{\tt R3D}, are needed for the astronomical community.  Thus we
distribute the package freely.  It provides a common package for
reducing data from any kind of fiber-fed spectrograph in general and
IFUs in particular. We plan to extend it to other types of
IFUs (such as TIGER-based lensarrays or Image Slicers). 
Use of {\tt R3D} will help soften the learning curve for working with 
IFS data obtained with different IFUs.

\begin{acknowledgements}
 
  We thank Dr. N. Cardiel for his invaluable help in the developing and testing
  process of {\tt R3D}, and the polishing process of this article. we offered
  him to be co-author in this work, due to the amount of work that he did on
  it, although he kindly (and honestly) refused.  He is always an inspiration
  of what an astronomer should be. 
  
  We thank Dr. M. Roth for his kindly help in cleaning the text of this
  article.  We thank Dr. R. Gredel, director of Calar Alto, for supporting
  this work and providing me with the working and telescope time for finishing
  it. We thank the referee, Dr. A. Kelz, for the useful comments on the
  article. We also thank Dr. C.R. Benn and D. Coe who kindly revised the
  English style of this article, introducing valuable comments and
  corrections.

  We thank support from the Spanish Plan Nacional de Astronom\'\i a program
  PNAyA2005, of the Spanish Ministery of Education and Science and the Plan
  Andaluz de Investigaci\'on of Junta de Andaluc\'{\i}a as research group
  FQM322.

\end{acknowledgements}

\end{document}